\documentclass[11pt]{article}
\usepackage{amsfonts}
\usepackage{color}
\usepackage{graphicx}
\usepackage{authblk}
\usepackage{latexsym,amsmath}
\usepackage{amssymb,array}
\usepackage{enumerate}
\makeatletter \oddsidemargin 0in \evensidemargin 0in \textwidth 16cm
\begin{document}
\newcommand{\la}{\lambda}
\newcommand{\up}{\upsilon}
\newcommand{\Up}{\Upsilon}
\numberwithin{equation}{section}
\newtheorem{d1}{Definition}[section]
\newtheorem{r1}{Remark}[section]
\newtheorem{thm}{Theorem}[section]
\newtheorem{l1}{Lemma}[section]
\newtheorem{ex}{Example}[section]
\newtheorem{cx}{Counterexample}[section]
\newtheorem{cor}{Corollary}[section]
\title{\bf
Generalization of the pairwise stochastic precedence order to the sequence of random variables}
\author[a,b]{Maxim Finkelstein\footnote{Corresponding author, email: FinkelM@ufs.ac.za}}
\author[c]{Nil Kamal Hazra}
\affil[a]{Department of Mathematical Statistics and Actuarial Science, University of the Free State, 339 Bloemfontein 9300, South Africa}
\affil[b]{ITMO University, Saint Petersburg, Russia}
\affil[c]{Department of Mathematics, Indian Institute of Information Technology, Design and Manufacturing Kancheepuram, Chennai 600127, India}
\date{December, 2018}
%\date{}
\maketitle
\begin{abstract}
We discuss a new stochastic ordering for the sequence of independent random variables. It generalizes the stochastic precedence order that is defined for two random variables to the case $n>2$. All conventional stochastic orders are transitive, whereas the stochastic precedence order is not. Therefore, a new approach to compare the sequence of random variables had to be developed that resulted in the notion of the sequential precedence order. A sufficient condition for this order is derived and some examples are considered.
\end{abstract}
\section{Introduction}
Stochastic orders are pairwise comparisons between two random variables. Numerous stochastic orders had been described and widely used in the literature including the most popular in reliability applications the usual stochastic order, the hazard rate order and the likelihood ratio order. The encyclopedic information on stochastic orders and their properties can be found in Shaked and Shantikumar~\cite{ss}. For the sake of completeness, the definitions of orders that are used in our paper are given below.
\\\hspace*{0.2 in}Let us first introduce the basic notation to be used throughout the paper. For an absolutely continuous random variable $T$, we denote the probability density function (pdf) by $f_T(\cdot)$, the cumulative distribution function (cdf) by $F_T(\cdot)$, the hazard rate function by $\lambda_T(\cdot)$, and
the survival/reliability
 function by $\bar F_T(\cdot)$.
\begin{d1}
Let $T_1$ and $T_2$ be two absolutely continuous random variables with respective supports $(l_{1},u_{1})$ and $(l_2,u_2)$,
where $u_1$ and $u_2$ may be positive infinity, and $l_1$ and $l_2$ may be negative infinity.
Then, $T_2$ is said to be larger than $T_1$ in
\begin{enumerate}
\item [$(a)$] likelihood ratio (lr) order denoted as $T_1\leq_{lr}T_2$, if
$${f_{T_2}(t)}/{f_{T_1}(t)} \text{ is increasing in } t\in(l_1,u_1)\cup(l_2,u_2);$$
\item [$(b)$] hazard rate (hr) order denoted as $T_1\leq_{hr}T_2$, if $${\bar F_{T_2}(t)}/{\bar F_{T_1}(t)}\text{ is increasing in } t \in (-\infty,max(u_1,u_2));$$
 \item [$(c)$] usual stochastic (st) order denoted as $T_1\leq_{st}T_2$, if $$\bar F_{T_1}(t)\leq \bar F_{T_2}(t) \text{ for all }t\in~(-\infty,\infty).$$
\end{enumerate}
\end{d1}
Using pairwise comparisons, the sequence of $n$ independent random variables $T_i$, $i=1,2,\dots,n$ can be also ordered as
\begin{eqnarray}
T_1\leq T_2\leq \dots \leq T_n\label{eq1}
\end{eqnarray}
in a suitable stochastic sense. For all mentioned basic stochastic orders \eqref{eq1} is transitive meaning  that $T_i\leq T_j$, for all $1\leq i<j\leq n$.
\\\hspace*{0.2 in}We now define the stochastic precedence (SP) order (see, e.g., Boland et al.~\cite{bsc}, Finkelstein~\cite{f}, Montes and Montes~\cite{mm}, to name a few).
\begin{d1}\label{de2}
Let $T_1$ and $T_2$ be two nonnegative independent random variables. Then, $T_2$ is said to be larger than $T_1$ in stochastic precedence (SP) order, denoted as $T_2\geq_{sp}T_1$, if
\begin{eqnarray}\label{eq2}
P(T_2\geq T_1)> P(T_1\geq T_2).
\end{eqnarray}
We write $T_1=_{sp} T_2$, if $P(T_2\geq T_1)=P(T_1\geq T_2)$.
\end{d1}
For continuous random variables, \eqref{eq2} can be equivalently written as
\begin{eqnarray}\label{eq3}
P(T_2\geq T_1)\geq 0.5.
\end{eqnarray}
This order is relevant in numerous engineering applications when e.g., stress-strength (Finkelstein~\cite{f}) or peak over the threshold probabilities are considered. The idea of using \eqref{eq3} as a reasonable tool for comparing random variables probably goes to Savage (\cite{s}, p. 245). The SP order can be appropriate in some problems as it directly describes probabilities of interest (distinct from other popular stochastic orders). It can be easily shown that the stochastic precedence order for independent random variables follows from the usual stochastic order. Thus, it is weaker and more flexible and can describe random variables with crossing reliability functions.
\\\hspace*{0.2 in}However, if we want to order the sequence in \eqref{eq1} with respect to the SP order, i.e.,
 \begin{eqnarray}\label{e00}
 T_1\leq_{sp}T_2\leq_{sp}\dots\leq_{sp}T_n,
 \end{eqnarray}
 then, non-necessarily, $T_i\leq_{sp}T_j$, $1\leq i<j\leq n$ meaning that this order is non-transitive (Santis et al.~\cite{sfs}). Let us call \eqref{e00}, for convenience, the chain stochastic precedence (CSP) order.
\begin{ex}
For simplicity of illustration of the non-transitivity property, consider the case of three discrete random variables with the following distributions (Blyth~\cite{b72})
 \begin{eqnarray*}
 &&P(T_1=3)=1, \\&&P(T_2=1)=0.4,\;
 P(T_2=4)=0.6, \\&&P(T_3=2)=0.6,\; P(T_3=5)=0.4.
 \end{eqnarray*}
Then, obviously,
$$P(T_1<T_2)=0.6,\;P(T_2<T_3)=0.64, \text{ and }P(T_3<T_1)=0.6.$$
Hence, $T_i\nleq_{sp}T_j$, $1\leq i<j\leq 3$.$\hfill\Box$
\end{ex}
\hspace*{0.2 in}Sometimes this non-transitivity for three random variables  is called  a voting paradox (Blyth~\cite{b72}).
Thus, the SP order in general can be non-applicable for ordering  sequences of random variables. However, it can be generalized to the transitive case on the basis of  Definition~\ref{de2} that compares probabilities of the corresponding events.
\begin{d1}\label{de3}
The sequence of $n$ independent random variables $T_i$, $i=1,2,\dots,n$ is ordered in the sense of the stochastic sequential precedence (SSP) order if it gives the maximal probability, e.g., to the event $T_1\leq T_2\leq \dots\leq T_n$, as compared with probabilities of events for all other permutations in the sequence of events $\{\mbox{\boldmath$J$}\}$, i.e.,
 \begin{eqnarray}\label{eq4}
 P_{1,2,\dots,n}\equiv P(T_1\leq T_2\leq \dots\leq T_n)\geq P_{\left\{\mbox{\boldmath$J$}\right\}},
 \end{eqnarray}
whereas the corresponding notation will be
 $$\left(T_1\leq T_2\leq \dots\leq T_n\right)_{ssp}.$$
\end{d1}
\hspace*{0.2 in}It is clear that for $n=2$, \eqref{eq4} reduces to \eqref{eq2}. We shall not be concerned that the absolute values in \eqref{eq4} can be very small, as we are interested in comparisons.
 \\\hspace*{0.2 in}We had just outlined this new order in our recent paper (Finkelstein et al.~\cite{fhc}) while considering the problem of obtaining the optimal sequence of activation of components in the warm standby system. We shall briefly refer to this meaningful example and then study some initial properties of the SSP order.
\begin{ex}
\emph{Warm standby system}. Consider $1$-out-of-$n$ warm standby system when one of the components is activated at $t=0$ (full load) and others are in the warm standby mode (reduced load). When the activated component fails, one of the operable standby components is activated, etc. The problem is to find the optimal activation sequence that maximizes the lifetime of the system in a suitable probabilistic sense. This open (for a general case) problem was solved in Finkelstein et al.~\cite{fhc}, where it was proved that if the lifetimes of the components are ordered in the SSP sense, then this sequence of activation (starting with the shortest lifetime) results in a system's lifetime that is larger than a lifetime of a warm standby system for any other sequence of activation in the SP order sense (see also 
%Hazra and Nanada~\cite{hn} and 
Zhai et al.~\cite{zypz}).
     \\\hspace*{0.2 in}An important feature of the developed approach is that it was shown that the ordering of the corresponding independent realizations of components' lifetimes, i.e,
        \begin{eqnarray}\label{eq5}
        t_1\leq t_2\leq \dots\leq t_n
        \end{eqnarray}
for the considered system, maximizes realization of its lifetime, i.e.  $s_{1,2,\dots,n}\geq s_{\left\{\mbox{\boldmath$J$}\right\}}$, where $s_{\left\{\mbox{\boldmath$J$}\right\}}$ denotes this realization for the sequence of activation $\left\{\mbox{\boldmath$J$}\right\}$, whereas the corresponding lifetimes are denoted by $S_{1,2,\dots,n}$ and $S_{\left\{\mbox{\boldmath$J$}\right\}}$, respectively. It follows from \eqref{eq4} that
 \begin{eqnarray}\label{eq6}
 P\left(S_{1,2,\dots,n}\geq S_{\left\{\mbox{\boldmath$J$}\right\}}\right)=\frac{P_{1,2,\dots,n}}{P_{1,2,\dots,n}+P_{\left\{\mbox{\boldmath$J$}\right\}}}\geq 0.5,
 \end{eqnarray}
which is the SP order.
 \\\hspace*{0.2 in}Thus, ordering of realization \eqref{eq5} results in the maximal realization of the lifetime of the system. However, the corresponding event has the maximal probability due to assumption \eqref{eq4}. Finally, \eqref{eq6} defines the SP order for system's lifetimes.$\hfill\Box$
\end{ex}
\hspace*{0.2 in}The reasoning in this example prompts us that a similar logic can be followed for some optimization problems, where the corresponding results for realizations of relevant random variables can be derived. For instance, as in the following example.
\begin{ex}
\emph{Coherent system}. Assume that we have a coherent system (Barlow and Proschan \cite{bp}) of $n$ independent components with lifetimes $T_i$, $i=1,2,\dots,n$. Let their realizations be ordered as in \eqref{eq5}. Assume that based on the structure of the system we know how to allocate these realizations to n `slots' of a system in order to maximize the realization of the system's lifetime. For instance, for $n=3$, we have: $t_1\leq t_2\leq t_3$. Let the system be a series-parallel with one component in series with the parallel structure of two components. Then, if the components lifetimes are ordered as SSP,
 \begin{eqnarray*}
 P_{1,2,3}\equiv P(T_1\leq T_2\leq T_3)\geq P_{\left\{\mbox{\boldmath$J$}\right\}},
 \end{eqnarray*}
and we use the largest lifetime for the series part ($T_3$) and the other two, for the parallel part, the system lifetime, similar to \eqref{eq6}, will be the largest in the SP order as compared with other permutations. However, it is also obvious that in this simple case, if the components are ordered in the sense of the usual stochastic order, this variant of allocation will be also the best in the same sense. Note that, as the lifetimes of the system for different variants of allocation are statistically dependent, we cannot say now that the usual stochastic order implies the SP order, which is true for the independent random variables. Thus, we just have different ordering for the components and for the variants of the system as compared with the usual stochastic order. The latter combination of the SSP order with the SP order could be more practically sound in various reliability comparison problems than comparisons based on the usual stochastic order.
\end{ex}
\section{Some properties of the SSP order}
While considering different stochastic orders for the pair of independent random variables, we are often interested in the relationships between them. It is well-known that for the simplest stochastic orders for two independent random variables, we have the following chain;
 $$LR \implies HR \implies ST \implies SP.$$
It is interesting to obtain some relationships between the SSP order for n independent random variables and other orders for this sequence. This topic needs further investigation. Below we present some initial results. We start with the following example that will help us to formulate the sufficient condition for the SSP order.
\begin{ex}
Let $n=3$ and $(T_1\leq T_2\leq T_3)_{ssp}$.
%Denote the corresponding cumulative distribution functions and probability density functions of these random variables by $F_i(\cdot)$, $f_i(\cdot)$, $i=1,2,3,$ respectively.
 There are six permutations. Let us consider the permutation $(1\;2\;3)$ and compare it with permutation $(1 \;3 \;2)$. For $P(T_1\leq T_2\leq T_3)\geq P(T_1\leq T_3\leq T_2)$ to hold, the following should hold:
 $$\int\limits_{x=0}^\infty \int\limits_{y=x}^\infty f_{T_1}(x)f_{T_2}(y)(1-F_{T_3}(y))dydx\geq \int\limits_{x=0}^\infty \int\limits_{y=x}^\infty f_{T_1}(x)f_{T_3}(y)(1-F_{T_2}(y))dydx.$$
For this, it suffices to show that
$$f_{T_2}(y)(1-F_{T_3}(y))\geq f_{T_3}(y)(1-F_{T_2}(y)).$$
However, the last inequality means that $\lambda_{T_2}(y)\geq \lambda_{T_3}(y)$, which, obviously, defines the corresponding hazard rate order, i.e.,
 $T_2\leq_{hr}T_3.$ $\hfill\Box$
 \end{ex}
Thus, the first guess would be that the hazard rate order is the sufficient condition for the SSP order. However, considering comparisons with other permutation does not support this and, in fact, the likelihood ratio ordering becomes the corresponding sufficient condition. This will be proved now in what follows for the general case.
We begin with the following lemma.
\begin{l1}\label{le2}
Let $\{T_i\}_{i=1}^n$ be a sequence of independent random variables such that $T_1\leq_{lr}T_2\leq_{lr}\dots\leq_{lr}T_n$. Further, let $\{i_1,i_2,\dots,i_j,\dots,i_k,\dots,i_n\}$ be a permutation of $\{1,2,\dots,n\}$. Then, for $1\leq i_j<i_k\leq n$,
\begin{eqnarray*}
&&P\left(T_{i_1}\leq \dots \leq T_{i_{j-1}}\leq T_{i_j}\leq T_{i_{j+1}}\leq \dots\leq T_{i_{k-1}}\leq T_{i_k}\leq T_{i_{k+1}}\leq\dots\leq T_{i_n}\right)
\\&&\geq P\left(T_{i_1}\leq \dots \leq T_{i_{j-1}}\leq T_{i_k}\leq T_{i_{j+1}}\leq \dots\leq T_{i_{k-1}}\leq T_{i_j}\leq T_{i_{k+1}}\leq\dots\leq T_{i_n}\right).
\end{eqnarray*}
{\bf Proof:} Note that
\begin{eqnarray}
&&P\left(T_{i_1}\leq \dots \leq T_{i_{j-1}}\leq T_{i_j}\leq T_{i_{j+1}}\leq \dots\leq T_{i_{k-1}}\leq T_{i_k}\leq T_{i_{k+1}}\leq\dots\leq T_{i_n}\right)\label{e1}
\\&&=\int\limits_{t_1=0}^\infty\dots\int\limits_{t_{j-1}=t_{j-2}}^\infty\int\limits_{t_j=t_{j-1}}^\infty\int\limits_{t_{j+1}=t_{j}}^\infty\dots\int\limits_{t_{k-1}=t_{k-2}}^\infty\int\limits_{t_k=t_{k-1}}^\infty\int\limits_{t_{k+1}=t_{k}}^\infty\dots \int\limits_{t_{n}=t_{n-1}}^\infty A(t_1,t_2,\dots, t_{n})dz,\nonumber
\end{eqnarray}
where
$$A(t_1,t_2,\dots,t_{n})=f_{T_{i_j}}(t_j)f_{T_{i_k}}(t_k)\left(\prod\limits_{r=1}^{j-1}f_{T_{i_r}}(t_r)\right)\left(\prod\limits_{s=j+1}^{k-1}f_{T_{i_s}}(t_s)\right)\left(\prod\limits_{u=k+1}^{n}f_{T_{i_u}}(t_u)\right),$$ and
$$dz=dt_{n}\dots dt_{k+1}dt_{k}dt_{k-1}\dots dt_{j+1}dt_j dt_{j-1}\dots dt_1,$$
and $$t_1\leq \dots\leq t_{j-1}\leq t_j\leq t_{j+1}\leq \dots \leq t_{k-1}\leq t_k \leq t_{k+1}\leq \dots\leq t_{n}.$$
Similarly,
\begin{eqnarray*}
&&P\left(T_{i_1}\leq \dots \leq T_{i_{j-1}}\leq T_{i_k}\leq T_{i_{j+1}}\leq \dots\leq T_{i_{k-1}}\leq T_{i_j}\leq T_{i_{k+1}}\leq\dots\leq T_{i_n}\right)
\\&&=\int\limits_{t_1=0}^\infty\dots\int\limits_{t_{j-1}=t_{j-2}}^\infty\int\limits_{t_k=t_{j-1}}^\infty\int\limits_{t_{j+1}=t_{k}}^\infty\dots\int\limits_{t_{k-1}=t_{k-2}}^\infty\int\limits_{t_j=t_{k-1}}^\infty\int\limits_{t_{k+1}=t_{j}}^\infty\dots \int\limits_{t_{n}=t_{n-1}}^\infty A(t_1,t_2,\dots,t_{n})dw,
\end{eqnarray*}
where
$$dw=dt_{n}\dots dt_{k+1}dt_{j}dt_{k-1}\dots dt_{j+1}dt_k dt_{j-1}\dots dt_1,$$and
 $$t_1\leq \dots\leq t_{j-1}\leq t_k\leq t_{j+1}\leq \dots \leq t_{k-1}\leq t_j \leq t_{k+1}\leq \dots\leq t_{n}.$$
 Since $t_j$ and $t_k$ are dummy variables, we interchange them in the above probability expression. Then, we get
 \begin{eqnarray}
&&P\left(T_{i_1}\leq \dots \leq T_{i_{j-1}}\leq T_{i_k}\leq T_{i_{j+1}}\leq \dots\leq T_{i_{k-1}}\leq T_{i_j}\leq T_{i_{k+1}}\leq\dots\leq T_{i_n}\right)\label{e2}
\\&&=\int\limits_{t_1=0}^\infty\dots\int\limits_{t_{j-1}=t_{j-2}}^\infty\int\limits_{t_j=t_{j-1}}^\infty\int\limits_{t_{j+1}=t_{j}}^\infty\dots\int\limits_{t_{k-1}=t_{k-2}}^\infty\int\limits_{t_k=t_{k-1}}^\infty\int\limits_{t_{k+1}=t_{k}}^\infty\dots \int\limits_{t_{n}=t_{n-1}}^\infty B(t_1,t_2,\dots,t_{n})dz,\nonumber
\end{eqnarray}
where
$$B(t_1,t_2,\dots,t_{n})=f_{T_{i_j}}(t_k)f_{T_{i_k}}(t_j)\left(\prod\limits_{r=1}^{j-1}f_{T_{i_r}}(t_r)\right)\left(\prod\limits_{s=j+1}^{k-1}f_{T_{i_s}}(t_s)\right)\left(\prod\limits_{u=k+1}^{n}f_{T_{i_u}}(t_u)\right),$$ and
 $$t_1\leq \dots\leq t_{j-1}\leq t_j\leq t_{j+1}\leq \dots \leq t_{k-1}\leq t_k \leq t_{k+1}\leq \dots\leq t_{n}.$$
 Further,
 \begin{eqnarray}
 A(t_1,t_2,\dots,t_{n})-B(t_1,t_2,\dots,t_{n})\geq 0\label{e3}
 \end{eqnarray}
  holds if, for $t_j\leq t_k$,
 $$f_{T_{i_j}}(t_j)f_{T_{i_k}}(t_k)\geq f_{T_{i_j}}(t_k)f_{T_{i_k}}(t_j,),$$
 or equivalently,
 $$T_{i_j}\leq_{lr} T_{i_k}, \text{ for }1\leq i_j<i_k\leq n,$$
 which follows from the hypothesis that $T_1\leq_{lr}T_2\leq_{lr}\dots\leq_{lr}T_n$. Thus, on using \eqref{e3}, the result follows from the expressions given in \eqref{e1} and \eqref{e2}.$\hfill\Box$
\end{l1}
\begin{thm}
Let $\{T_i\}_{i=1}^n$ be a sequence of independent random variables such that $T_1\leq_{lr}T_2\leq_{lr}\dots\leq_{lr}T_n$. Then $(T_1\leq T_2\leq\dots T_n)_{ssp}$.
\end{thm}
{\bf Proof:} Note that a set of $n$ random variables could be arranged in $n!$ different ways by interchanging any two of them. Thus, the proof follows from repetitive use of Lemma~\ref{le2}. For instance, let us consider $n=3$. Then, from Lemma~\ref{le2}, we have
\begin{eqnarray}\label{eq7}
P(T_1\leq T_2\leq T_3)\geq P(T_1\leq T_3\leq T_2)\geq P(T_3\leq T_1\leq T_2).
\end{eqnarray}
Again,
\begin{eqnarray}\label{eq8}
P(T_1\leq T_2\leq T_3)\geq P(T_3\leq T_2\leq T_1).
\end{eqnarray}
and
\begin{eqnarray}\label{eq9}
P(T_1\leq T_2\leq T_3)\geq P(T_2\leq T_1\leq T_3)\geq P(T_2\leq T_3\leq T_1).
\end{eqnarray}
On using \eqref{eq7}, \eqref{eq8} and \eqref{eq9}, we get
$(T_1\leq T_2\leq T_3)_{ssp}$.$\hfill\Box$
\\\hspace*{0.2 in}In the following theorem, we show that the SSP order is stronger than the SP order for a sequence of $n$ independent random variables. 
\begin{thm}
Let $\{T_i\}_{i=1}^n$ be a sequence of independent random variables such that $(T_1\leq T_2\leq\dots \leq T_n)_{ssp}$. Then $T_i\leq_{sp}T_j$, for all $1\leq i<j\leq n$.
\end{thm}
{\bf Proof: }Let $N$  be a `sufficiently large' number of trials for the sequence $\{T_i\}_{i=1}^n$, whereas $N_{1,2,\dots,n}$ denote the number of realizations (out of $N$), that result in \eqref{eq5}.
 \\\hspace*{0.2 in}Select $T_i$ and $T_j$, where $1\leq i<j\leq n$. Inequalities \eqref{eq5} correspond to the case when their realizations are ordered as $t_i\leq t_j$. Now we consider realizations where $t_i\geq t_j$ with all other realizations of other random variables being the same as for the previous case. Denote the overall number of realizations of the latter kind by $N_{1,2,\dots,i-1,j,i+1,\dots, j-1,i,j+1,\dots,n}$. From our assumption $(T_1\leq T_2\leq\dots \leq T_n)_{ssp}$, it follows that
 $$N_{1,2,\dots,i-1,j,i+1,\dots, j-1,i,j+1,\dots,n}\leq N_{1,2,\dots,n}.$$
But this means that
 $$N_{j,i}\leq N_{i,j},$$
where $N_{i,j}$ and $N_{j,i}$ are the numbers of realizations for the pair of random variables $T_i$ and $T_j$ for which $T_i\leq T_j$ and $T_i\geq T_j$, respectively. But this, in fact, means the SP order for this pair by definition. Thus:
$T_i\leq_{sp}T_j \text{ for all }1\leq i<j\leq n.$ $\hfill\Box$
\\\hspace*{0.2 in}As an immediate consequence of the above theorem, we have the following corollary, which shows that the SSP order implies the CSP order that already obeys the transitivity property.
\begin{cor}
If $(T_1\leq T_2\leq\dots T_n)_{ssp}$, then $T_1\leq_{sp} T_2\leq_{sp}\dots \leq_{sp}T_n$.
\end{cor}
%Let us call (7), for convenience, the chain stochastic precedence (CSP) order. Note that, in general, it can be non-transitive.
%Thus, we had arrived at the following result.
\section{Concluding remarks}
The stochastic precedence order is natural in various engineering applications when, e.g., stress-strength or peak over the threshold probabilities are considered. It can be attractive for probabilistic description of real-life problems as it directly describes probabilities of interest (distinct from other popular stochastic orders). It is well known that the usual stochastic order implies the SP order, which gives the corresponding sufficient condition.
    \\\hspace*{0.2 in} However, distinct from the conventional stochastic orders (e.g., the usual stochastic order, the hazard rate order and the likelihood ratio order) that are transitive when ordering the sequence of random variables, the SP order in this case, can be non-transitive.
     \\\hspace*{0.2 in}Therefore, in this note, we discuss the new stochastic ordering for  the sequence of independent random variables that is called the sequential stochastic precedence (SSP) order. It generalizes the stochastic precedence order that is defined for two random variables to the case $n>2$ and by definition is transitive.
    \\\hspace*{0.2 in}We show that the likelihood ratio ordering is the sufficient condition for the SSP ordering of the sequence of the independent random variables. Moreover, the SSP order implies the SP order in this sequence.
\subsection*{Acknowledgements}
\hspace*{0.2 in}The second author sincerely acknowledges the financial support from the IIITDM Kancheepuram, Chennai.
 
\end{document}